\documentclass{emulateapj}
\usepackage{amssymb}
\begin{document}
\title{Outflow, Infall and Protostars in the Star Forming Core W3-SE}
\author{Lei Zhu\altaffilmark{1,2}, Jun-Hui Zhao\altaffilmark{2} \&
M. C. H. Wright\altaffilmark{3}}
\altaffiltext{1}{Department of Astronomy, Peking University}
\altaffiltext{2}{Harvard-Smithsonian
Center for Astrophysics, 60 Garden Street, Cambridge, MA 02138;
lzhu@cfa.harvard.edu, jzhao@cfa.harvard.edu}
\altaffiltext{3}{Department of Astronomy, University of California,
Berkeley, Berkeley, CA 94720}

\begin{abstract}
We report new results on outflow and infall in the star forming cores
W3-SE SMA-1 and SMA-2 based on analysis of $\sim$2.5$\arcsec$
resolution observations of the molecular lines HCN(3-2), HCO$^+$(3-2),
N$_2$H$^+$(3-2) and CH$_3$OH(5$_{2,3}$-4$_{1,3}$) with the
Submillimeter Array. A high-velocity bipolar outflow originating from
the proto-stellar core SMA-1 was observed in the HCN(3-2) line, with
a projected outflow axis in position angle 48$\arcdeg$. The detection
of the outflow is confirmed from other molecular lines. An inverse
P-Cygni profile in the HCN(3-2) line toward SMA-1 suggests that at
least one of the double cores accretes matters from the molecular
core. A filamentary structure in the molecular gas surrounds SMA-1
and SMA-2. Based on the SMA observations, our analysis suggests that
the double pre-stellar cores SMA-1 and SMA-2 result from
fragmentation in the collapsing massive molecular core W3-SE, and
it is likely that they are forming intermediate to high-mass stars
which will be new members of a star cluster in the W3-SE region.
\end{abstract}

\keywords{ISM: clouds -- ISM: kinematics and dynamics -- ISM: molecules
-- ISM: jets and outflows -- radio lines: ISM -- stars: formation}

\section{Introduction}

One of the key questions in the study of star formation is how
protostars accrete material from their parent molecular clouds.
Accretion around protostars with intermediate and high mass is not
well understood due to lack of high-resolution observations.
Observation of infall is needed to provide direct evidence for
accretion. Low resolution observations of a blue-dominated
double-peaked spectral profile as expected from a model of
self-absorption in infalling gas clumps
\citep[NGC 1333-IRAS 2 as an example; see][]{wtho96,wtho01,jorg04a,pere06}
do not unambiguously identify infall because massive cores usually
contain bipolar outflows, disk rotation and multiple sub-cores in
addition to gas infall. Complex dynamical processes can produce
characteristics similar to those expected from the self-absorption
model of a molecular core with infall. It is crucial to carry out
high-angular and spectral resolution observations along with
comprehensive modeling of the dynamical processes involved in a
massive star forming region.

There is growing evidence that massive stars form in groups,
but accretion modeling has been for material concentrated in
monolithic cores \citep{mcke07}. The formation of binary and multiple
stellar systems, especially those with two comparable-mass protostars
and a large separation, is poorly understood \citep{mcke07}. Detailed
high-resolution observations of a binary/multiple proto-stellar system
will provide valuable data for understanding accretion and
outflow from each of the sub-cores in the system.

W3-SE is a dense molecular core located SE of the well-known massive
star formation region W3-Main. We adopted a distance of $D$ = 2 kpc for
W3-SE \citep{blit82,xu06}. The spectral energy distribution (SED),
along with a double continuum core
revealed in our previous paper \citep{zhu10} suggests that a total mass
65~M$_\sun$ for the W3-SE core over a scale of 10$\arcsec$ is mainly
composed of two proto-stellar cores of comparable masses, which
are separated by 3.5$\arcsec$
or 7000 AU as revealed in the continuum observations with the
Submillimeter Array (SMA).\footnote{The Submillimeter
Array is a joint project between the Smithsonian Astrophysical
Observatory and the Academia Sinica Institute of Astronomy and
Astrophysics and is funded by the Smithsonian Institution and the
Academia Sinica.}
Observations of HCO$^+$(1-0) using the CARMA telescope at
6$\arcsec$ resolution show that outflows appear to dominate
the observed spectra although infall might also contribute to the
spectrum toward the main molecular core which coincides with the
double cores. Spectral-line observations with higher angular
resolution are necessary to separate the contributions of
infall and outflow from the double cores in W3-SE and to investigate
the possibility of a protobinary system.

In this paper, we present results of analysis of spectral-line data
observed with the SMA at $\sim$2.5$\arcsec$ resolution.
Section 2 describes the observations and data reductions. Section 3
shows the results on the basis of our analysis of the molecular line
data including HCN(3-2), HCO$^+$(3-2), N$_2$H$^+$(3-2) and
CH$_3$OH(5$_{2,3}$-4$_{1,3}$). In Section 4, we present an
interpretation of the observations in the context of outflow and
infall. Section 5 gives a summary and conclusions.

\section{Observations}

The observations toward W3-SE consist of two tracks obtained at
$\lambda$~1.1 mm using the SMA in the compact configuration with an
angular resolution of $\sim$2.5$\arcsec$ in 2008 October 27th and
December 20th. The observations were centered at R.A.(J2000) = 02$^{\rm
h}$25$^{\rm m}$54$^{\rm s}$.50, DEC(J2000) =
62$\arcdeg$04$\arcmin$11$\arcsec$.0, which is used as the reference
position in the following analysis. The LO frequency ($\nu_{\rm LO}$)
for the 2008 October track was tuned to 271.7 GHz and $\nu_{\rm LO}$ =
285.3 GHz for the other. The observations consist of a lower side band
(LSB) and an upper side band (USB) separated by 10 GHz, with a
bandwidth of 2 GHz for each sideband. The typical system temperatures
(T$_{\rm sys}$) are 180-200 K and 200-250 K for the observations in
October and December, 2008, respectively.

Data reduction and imaging were carried out in {\it
MIRIAD} \citep{saul95} with specific implementations for SMA data
reduction\footnote{http://www.cfa.harvard.edu/sma/miriad}. The bandpass
was calibrated using QSOs 3C454.3 for the lower-frequency track
and 3C273 for the higher-frequency track. For both tracks the complex
gains were calibrated using QSO 0244+624. The flux density scale was
bootstrapped from Uranus using the SMA planet model. By combining the
visibility data from both LSB and USB, the continuum images achieve rms
noises of 1.5 mJy and 3.0 mJy for the data observed in 2008 October and
December, respectively. Spectral line images were made by subtracting
the continuum emission from each visibility data using the {\it MIRIAD}
task {\it UVLIN}.

\section{Results}

\subsection{HCN(3-2)}

A transition of hydrogen cyanide, HCN(3-2) at $\nu_0$ = 265.886 GHz,
was observed in the LSB of the 2008 October observations. Figure 1a
presents the HCN(3-2) channel maps, showing multiple velocity
components distributed in the W3-SE core. In the velocity range between
--39.3 and --38.0 km s$^{-1}$ a gap with no significant line emission
signals is observed. The figure also shows compact HCN(3-2) components
in the high-velocity channels ($<$ --46.7 km s$^{-1}$ and $>$ --30.5
km~s$^{-1}$). Figure 2a shows the integrated line intensity map of
HCN(3-2), suggesting the presence of three emission peaks from the
HCN(3-2) line within the main molecular emission component A, as
observed by CARMA at an angular resolution of 6$\arcsec$
\citep{zhu10}. We denote the three sub-components as A-NE, A-SW and
A-SE, respectively. The HCN(3-2) line components A-NE and A-SW are
located 2$\arcsec$ northeast and 4$\arcsec$ southwest of the
proto-stellar core SMA-1 while the component A-SE nearly coincides
with SMA-2. The integrated intensity of HCN(3-2) line emission in A-NE
and A-SW is stronger than that from A-SE.

Figure 3 shows the spectra of the HCN(3-2) line in the core region of
W3-SE with a boundary outlined by the square in Figure 2a. A significant
absorption feature is shown in the core region at velocities in the
range between --39.3 and --38.0 km s$^{-1}$. The component A-NE
appears to be associated with a broad blue-shifted line wing, while
A-SW shows a broad red-shifted line wing. The highly blue-shifted line
wing associated with A-NE shows a maximum
absolute radial velocity of 35 km s$^{-1}$ (above the 5 $\sigma$
cutoff), with respect to SMA-1. Here an LSR systemic velocity $V_{\rm sys}$ =
--39.1 km s$^{-1}$ is used, which is determined from the peak velocity of
the optically thin line CH$_3$OH(5$_{2,3}$-4$_{1,3}$). In the region of
A-SW, the red-shifted line wing shows a maximum absolute radial
velocity of 33 km s$^{-1}$ (above 5 $\sigma$) with respect to SMA-1.
The velocity profile for both blue- and red-shifted broad line wings
can be fitted to a power law ($\sim$$|V-V_{\rm
sys}|$$^{-\gamma}$) with an index of $\gamma\approx2.0$. The power-law
distribution of molecular content as function of velocity observed is
consistent with  numerical models of bow-shock outflow
\citep[e.g.,][]{down99,lee01}.

In Figure 4a, we show the high-velocity gas in both the blue-shifted
(--74.1 to --45.2 km s$^{-1}$) and red-shifted (--31.9 to --5.8 km
s$^{-1}$) outflow lobes, located NE and SW of the
proto-stellar core SMA-1 with separations of 2.2$\arcsec$ and
5.0$\arcsec$, respectively. The configuration of the
emission from the high-velocity wings, along with the location of SMA-1
suggests a bipolar outflow originating from SMA-1, a possible energy
source in the region. The major axis of the outflow is shown as a
straight line connecting the two peaks of the blue- and red-shifted
knots in Figure 4a, at P.A. = 48$^\circ$.
Although the location of SMA-1 appears to be offset
from the outflow axis by $\sim$0.4$\arcsec$, about 5\% of the
distance between the two high-velocity knots, such an apparent offset
could be due to the uncertainty ($\sim$0.2$\arcsec$) in the peak
position of SMA-1 given by Gaussian fitting.

In summary, the high-velocity HCN(3-2) line emission from the two
compact molecular components A-NE and A-SW appears to be excited by a
high-velocity bipolar outflow originating from the proto-stellar core
SMA-1.

\subsection{HCO$^+$(3-2)}

The transition of HCO$^+$(3-2) at 267.558 GHz was also included in the
LSB of the observations at $\nu_{\rm LO}$ = 271.7 GHz. Figure 1b shows
the channel maps of the HCO$^+$(3-2) line. The overall distribution of
HCO$^+$ gas in velocity is consistent with that of HCO$^+$(1-0)
observed with CARMA \citep{zhu10} and similar to that of HCN(3-2) in
the SMA observations, showing multiple emission peaks at various
velocities and a gap in the velocity range between --38.9 and --37.9 km
s$^{-1}$.

However, not all of the emission peaks observed in HCO$^+$(3-2)
coincide with those in HCN(3-2). In the integrated intensity map of
HCO$^+$(3-2) in Figure 2b, the emission peak A-NE observed in HCN(3-2)
appears to be embedded in the surrounding emission, and A-SW in
HCO$^+$(3-2) is not as prominent as it is in HCN(3-2). Instead,  strong
emission in HCO$^+$(3-2) appears to be concentrated in a bright ridge
0.5$\arcsec$ northeast of the line connecting the double proto-stellar
cores SMA-1 and SMA-2. Outside the main core A, the molecular sub-core
B and the molecular tail C are also present, in good agreement with the
image of HCO$^+$(1-0) observed with CARMA. In comparison with the HCN(3-2)
image, the HCO$^+$(3-2) line appears to trace a more extended envelope
around the double proto-stellar cores.

As Figure 5 shows, the observed HCO$^+$(3-2) spectra are also found
with broad line wings at the position of SMA-1 and its adjacent region,
although the line wings in HCO$^+$(3-2) are not as significant and broad
as that in HCN(3-2) spectra. The integrated intensity map for the blue-
and red-shifted line wings is shown in Figure 6a. The velocity ranges for
the integrations are (--57.9, --45.3) km s$^{-1}$ and (--31.8, -18.8)
km s$^{-1}$, with the same lower velocity limits as that for HCN(3-2).
In Figure 6a, the compact blue- and red-shifted lobes are offset from
SMA-1 similar to the case of HCN(3-2), although they are closer to SMA-1
(1.3$\arcsec$ for the blue lobe, and 3.4$\arcsec$
for red lobe) than the lobes in HCN(3-2). This could be explained if
HCN(3-2) emission is excited close to bow shocks, while HCO$^+$(3-2)
traces colder and more extended gas which is farther from bow
shocks. Another possible reason is that the emission of the high-velocity
outflow lobes might be blended by the HCO$^+$(3-2) emission from a
 large-scale outflow \citep{zhu10} and other dynamical
components associated with SMA-2. Indeed, in Figure 6a a red-shifted
HCO$^+$(3-2) component is found close to SMA-2 at  high velocity.
The same component is also marginally detected in HCN(3-2) (see Figure 4a),
and it might be an outflow component associated with SMA-2.

The low-velocity line wings of the HCO$^+$(3-2)
spectra were also integrated with the velocity ranges (--44.9, --41.7)
km s$^{-1}$ and (--36.3, --32.3) km s$^{-1}$ for blue and red lobes,
respectively. The velocity ranges have similar lower velocity
limits as the ones used in previous HCO$^+$(1-0) study \citep{zhu10}.
As Figure 6b shows, the distribution of blue- and red-shifted low-velocity
HCO$^+$(3-2) gas is consistent with the results of HCO$^+$(1-0),
implying a possible large-scale outflow in a NW-SE direction.

\subsection{N$_2$H$^+$(3-2)}

A transition of protonated nitrogen, N$_2$H$^+$(3-2) at
$\nu_{0}$ = 279.512 GHz was included in the LSB of the 2008 December
observations. Figure 2c shows the integrated intensity map of
N$_2$H$^+$(3-2), indicating multiple N$_2$H$^+$(3-2) emission clumps
which form a curved filamentary emission region. Among the numerous
N$_2$H$^+$(3-2) emission peaks, the two strongest emission peaks appear
to be located near SMA-1 and SMA-2 with a displacement of
$\sim$0.5$\arcsec$ from the continuum peaks, suggesting that a close
relationship between the dense molecular gas and the proto-stellar
cores.

The bright N$_2$H$^+$(3-2) emission region is located SW of the
double continuum cores, opposite to the locations of the bright
emission ridge of HCO$^+$(3-2). Emission at the locations of SMA-1 and
SMA-2 might be influenced by excitation or chemical conditions. The
primary formation routes for both HCO$^+$ and N$_2$H$^+$ involve
reactions with H$_3^+$. For a higher CO abundance, the main formation
route is the reaction between CO and H$_3^+$ in forming HCO$^+$. While when
the CO abundance gets low enough, the reaction between H$_3^+$ and N$_2$
in forming N$_2$H$^+$ becomes the main mechanism for the removal of H$_3^+$
\citep{jorg04b}. In addition, as CO returns to the gas phase from a
freezing-out phase, destruction of N$_2$H$^+$ through reactions with CO
becomes the dominant removal mechanism for N$_2$H$^+$. The possible
anti-correlation of the bright emission regions between N$_2$H$^+$(3-2)
and HCO$^+$(3-2) near the proto-stellar cores might reflect a gradient
of the CO abundance across the W3-SE core.

The line profiles of N$_2$H$^+$(3-2) spectra towards SMA-1 and SMA-2
are shown in Figure 7. We noticed the skewness of the line profiles, {\it i.e.},
at the position of SMA-1, the N$_2$H$^+$(3-2) spectrum mimics a ``blue-profile",
while toward SMA-2 N$_2$H$^+$(3-2) is like a ``red-profile". The main
hyperfine structures of the transition $J$=3-2 for N$_2$H$^+$
spread over a frequency range of only a few 100 kHz (100 kHz $\sim$ 0.1 km s$^{-1}$),
while the velocity difference between the two emission peaks of the N$_2$H$^+$(3-2)
spectrum toward SMA-1 is $\sim$5.0 km s$^{-1}$. Therefore, it is unlikely that
the observed skewness in the  N$_2$H$^+$(3-2) line profile is due to
hyperfine structure.

In order to model the overall kinematics of the molecular filament in the region,
a centroid velocity map of N$_2$H$^+$(3-2) was made (see Figure 8a) by calculating
$v_{\rm cen} = \int{vI_v}{\rm d}v/\int{I_v}{\rm d}v$ at each position. Here $v_{\rm cen}$ is
the centroid velocity (intensity-weighted mean velocity), and $I_v$ is the
intensity of the spectrum at a given velocity $v$.
In general, the distribution of the centroid velocity shows a gradient of
30$\pm$10 km s$^{-1}$ pc$^{-1}$ over a spatial scale of $\sim$20$\arcsec$, i.e.,
the gas in NW is red-shifted while that in SE is blue-shifted. This velocity
structure appears to be similar to the large-scale velocity structure traced by
HCO$^+$(1-0) that is believed to be a result of outflow \citep{zhu10}.

\subsection{CH$_3$OH(5$_{2,3}$-4$_{1,3}$)}

The methanol transition CH$_3$OH(5$_{2,3}$-4$_{1,3}$) at
$\nu_0=266.838$ GHz was observed in the LSB in the October 2008
observations. Figure 2d shows the integrated intensity map of this
transition. The two main CH$_3$OH(5$_{2,3}$-4$_{1,3}$) emission peaks
coincide with the two proto-stellar cores SMA-1 and SMA-2 within the
uncertainty in position ($\sim$0.2$\arcsec$), indicating that
this line traces the densest part of the cores and could be used to
determine the systemic velocity. The methanol emission
appears to mainly arise from the two proto-stellar cores as well as the
high-velocity outflow from SMA-1. In addition, in region~B, a weak
signal of CH$_3$OH(5$_{2,3}$-4$_{1,3}$) is detected at a 3$\sigma$ level. No
significant emission from CH$_3$OH(5$_{2,3}$-4$_{1,3}$) was detected
towards region~C. The distribution of the centroid velocity in CH$_3$OH
emission (Figure 8b) has no significant velocity gradients, although at
the northwest boundary and the position of A-SW the centroid velocity
is slightly red-shifted.

\subsection{Comparison of Molecular Line Spectra towards SMA-1 and SMA-2}

Figure 7 shows the spectra from the peak positions of SMA-1 and SMA-2
in the lines of HCN(3-2), HCO$^+$(3-2), N$_2$H$^+$(3-2) and
CH$_3$OH(5$_{2,3}$-4$_{1,3}$), respectively. In general, double-peak
profiles appear to characterize the optically thick lines toward both
cores from different molecules. For a given molecular line, the
double-peaked profile of SMA-1 is dominated by the blue-shifted feature,
whereas in SMA-2, the red-shifted peak is stronger than the blue-shifted
one. For a given core, the separation between the
peaks of the double-peak line profile appears to vary. For example, in
the case of SMA-1, the velocity separations between the two spectral
peaks vary significantly from 7.6 km s$^{-1}$ for the HCN(3-2) line to
4.0 km s$^{-1}$ for both the HCO$^+$(3-2) and N$_2$H$^+$(3-2) lines,
suggesting that different molecular lines may trace different dynamic
processes or chemistry in a proto-stellar core. For  the HCN(3-2) line
from SMA-1, the large velocity separation of 7.6 km s$^{-1}$ between
the double emission peaks is probably due to the high-velocity bipolar
outflow because the HCN(3-2) line is the best tracer of the shocked and
compressed medium among the four observed molecular lines shown in
this paper.

In contrast to the double spectral peaks observed in HCN(3-2) and
HCO$^+$(3-2), the hot molecular line CH$_3$OH(5$_{2,3}$-4$_{1,3}$)
shows a single-peaked profile toward both the proto-stellar cores.
Gaussian fits to the spectra show that
CH$_3$OH(5$_{2,3}$-4$_{1,3}$)  are peaked at --39.1$\pm$0.1 km
s$^{-1}$ for both SMA-1 and SMA-2. We use this as the
systemic velocity in this paper, i.e. $V_{\rm sys}$ = --39.1 km s$^{-1}$.
The apparent peak velocities of the spectra
have a difference of $\sim$0.4 km s$^{-1}$, equivalent to the width of
one channel. Figure 8b also suggests a difference $\sim$0.5 km s$^{-1}$
between the centroid velocities of CH$_3$OH(5$_{2,3}$-4$_{1,3}$) toward
the two cores. However, this inconsistency could be due to the slightly
asymmetric CH$_3$OH(5$_{2,3}$-4$_{1,3}$) line profile toward SMA-2 and
contamination from  the complex kinematics in W3-SE region (e.g. the
outflow from SMA-1). Thus, we conclude that there is no significant
difference between the radial velocities of CH$_3$OH(5$_{2,3}$-4$_{1,3}$)
from SMA-1 and SMA-2.

An emission bump at --40.2 km s$^{-1}$ is present in the spectral dip
between the double HCN(3-2) emission peaks toward SMA-1 (see left-top
panel in Figure 7). The velocity of the emission bump appears to
correspond to the velocity of the blue-shifted peak in both the
HCO$^+$(3-2) and N$_2$H$^+$(3-2) lines, which is blue-shifted with
respect to the systemic velocity. In addition, the red-shifted dip
between the two main emission peaks appears to be an absorption feature
and the intensity becomes negative ($\sim$ 5 $\sigma$) at --37.6
km s$^{-1}$. We will discuss these spectral features in Section 4.2.1.

\section{Discussion}

\subsection{High-velocity Bow-shock Outflow in HCN(3-2)}

Two possible mechanisms for accelerating molecular gas in outflows from
proto-stellar cores are stellar winds and highly collimated jets
\citep[e.g.][]{down99}. The broad velocity distribution with a power
law in both blue- and red-shifted HCN(3-2) line wings and the
morphology of the high-velocity HCN(3-2) lobes suggests that
high-velocity HCN(3-2) emission is excited at jet-driven bow shocks. An
outflow model with jet-driven bow shocks has been studied both
analytically and numerically \citep[e.g.][]{down99,lee00}. In the case
of W3-SE, a P-V diagram along the major axis of the bipolar outflow
observed in the HCN(3-2) line (Figure 9) shows similar characteristics
to that predicted from the bow-shock model. The broad blue- and
red-shifted velocity gas components show significant offsets of
2.2$\pm$0.2$\arcsec$~and 5.0$\pm$0.2$\arcsec$ from SMA-1, providing
evidence that the high velocity HCN(3-2) gas is indeed excited and
accelerated at the bow-shocked sites outside of the proto-stellar core
SMA-1.

To characterize the observed high-velocity outflow, we adopted a simple
model of bow-shock outflow \citep{down99}, in which the geometry and
velocity of the molecular gas along the bow-shock surface can be worked
out in analytical forms. Using the model illustrated in Figure 10,
a cylindrical symmetric outflow radius ($r$) and major
axis ($z$) plane are used following \cite{down99}. We derived equations
for the four boundaries that outline the observed P-V
diagram (see Appendix). A power-law (${\displaystyle z\sim|r|^s }$)
with an index $s = 2$ for the locus of the bow-shock surface is assumed
in fitting the model to the observed P-V diagram of the bipolar outflow
from SMA-1 as discussed below. The systemic velocity, V$_{\rm sys}$ =
--39.1 km s$^{-1}$, and the projected angular offsets of the blue- and
red-shifted bow-shock tips ($a_B \approx 2.2\arcsec$ and $a_R \approx
5.0\arcsec$) from the outflow origin have been determined directly from
the peak velocity of the hot molecular line
CH$_3$OH(5$_{2,3}$-4$_{1,3}$) (Figure 7) and the HCN(3-2) outflow map
(Figure 4a), respectively. The remaining two free parameters $\alpha$
(inclination angle) and $v$ (the velocity of bow-shock with respect to
the origin of the outflow) were assumed to have equal values for the
blue- and red-shifted lobes, and they can be determined by comparing
the observed P-V diagram and the curve from the bow-shocked outflow model.
Figure 9 shows the observed P-V diagram in contours
and the curves derived from the bow-shock outflow model in thick lines.
A shock velocity $v$ = 40 km s$^{-1}$ and three different inclination
angles $\alpha$ = 55$\arcdeg$, 70$\arcdeg$, and 85$\arcdeg$ were used
to generate the curves. It appears that the model with $v$ = 40$\pm$10
km s$^{-1}$ and $\alpha$ = 70$\pm$5$\arcdeg$ best fits the overall line
wings of the P-V diagram. To estimate the uncertainty (1~$\sigma$) in
the values of $v$ and $\alpha$, we added small offset to cause the
derived model curve to offset by one contour interval of the P-V
diagram (3~$\sigma$). The inferred shock velocity of the bipolar outflow
implies a dynamic age of 3$\times$10$^3$ yr for the the proto-stellar
core SMA-1, which is an order of magnitude younger than that inferred
for the possible large-scale molecular outflow in the region \citep{zhu10},
indicating that the source(s) embedded in SMA-1 is probably very young.
On the other hand, SMA-1 is known to be associated with mid-IR source
\citep{zhu10} which implies the existence of heating source in SMA-1.
We offer two possible explanations, i.e. the protostar in SMA-1 has
a large mass and evolves more rapid than classic low-mass stars, or
there are multiple YSOs embedded in SMA-1 and the compact outflow and
the mid-IR source are actually associated with different YSOs.

\subsection{Infalling Gas toward SMA-1}

\subsubsection{Possible Infall Signature}

The blue-dominated double-peaked profile of the HCO$^+$(1-0)
spectrum toward core A observed using CARMA at 6$\arcsec$
resolution was fitted using a model with both outflow and infall
components \citep{zhu10}. The fit suggests that outflow significantly
contributes to the spectral profile, particularly in the
high-velocity wings. The high-velocity outflow lobes from SMA-1 are
separated from the proto-stellar core in the SMA HCN(3-2)
observations at (2.5$\arcsec$$\times$2.3$\arcsec$) resolution.
Contamination from the high-velocity outflow in the spectra toward
the cores has been reduced to a level where an inverse P-Cygni
profile, a signature of infall, would become noticeable if it exists.

The HCN(3-2) spectra (Figure 3), show blue-shifted emission at
--40.2 km s$^{-1}$ toward SMA-1 and the adjacent region, with
a sharp dip in the emission at --37.6 km s$^{-1}$. The systemic
velocity of --39.1 km s$^{-1}$ lies between the two spectral
features. The complexity of the HCN(3-2) line profile
toward SMA-1 might be due to a superimposition of an inverse
P-Cygni profile and a broader double-peaked profile from the
high-velocity bipolar outflow. Such an explanation is consistent
with the hypothesis that gas is infalling to SMA-1 \citep{zhu10}.

Figure 7 shows that toward SMA-1 the intensity of the HCN(3-2) dip
at --37.6 km s$^{-1}$ becomes negative. The spectral dip could be
partially caused by unsampled short-spacings. However, in our
observations the shortest projected baseline is 5.4 k$\lambda$,
corresponding to a spatial scale of $\sim$38$\arcsec$. Thus for
the spectrum toward SMA-1 with a beam size
(2$\arcsec$.5$\times$2$\arcsec$.3), flux missing should not be a
severe problem. We can roughly estimate the short spacing effect
using the previous study of CARMA data. For the HCO$^+$(1-0) spectra at
the peak position, the flux density of the spectral dip was raised to
0.5 Jy beam$^{-1}$ from 0.0 Jy beam$^{-1}$ after combining with
single-dish data. Assuming that the missing flux density of 0.5 Jy
beam$^{-1}$ is evenly distributed in a structure on a scale of the
CARMA beam (6$\arcsec$.3$\times$5$\arcsec$.7), and toward SMA-1 the
flux density of HCN(3-2) is the same as HCO$^+$(1-0), the missing flux
in HCN(3-2) at the velocity of the dip is less than 0.1 Jy within the
SMA beam. On the other hand, the intensity of HCN(3-2) towards
SMA-1 is $\sim$--1 K ($\sim$--0.4 Jy beam$^{-1}$) at the spectral dip.
Therefore, in the case of W3-SE the missing flux caused by unsampled
short spacings has only a small influence in the core region on
a scale of the beam size, and the observed negative intensity for
the HCN(3-2) spectral dip is likely to be real.

If the emission bump is explained as an inverse P-Cygni
feature, the infalling gas associated with SMA-1 appears to
be confined to a region $\sim$6$\arcsec$ ($\sim$12000 AU)
with velocities in a range of (--40 to --38 km s$^{-1}$)
that is close to the systemic velocity of the proto-stellar
core. We made an integrated intensity map of this emission
in a velocity range from --40.0 to --39.5 km s$^{-1}$
(Figure 4b). This image clearly shows that the HCN(3-2)
feature is associated with SMA-1, thus SMA-1 is likely to
be the gravitational center. Weak HCN(3-2) emission to
the SE of SMA-1 at a similar radial velocity, could blend
with the observed HCN(3-2) spectra from SMA-1.

\subsubsection{A General Model Including Both Infall and Outflow}

In order to separate the possible inverse P-Cygni (infall) feature
from the outflow, we adopted a characteristic model with a continuum
core in the middle comprised by two layers of infalling gas and two
layers of outflowing gas located outside of the infall core, which is
hereafter referred to as a four-layer model \citep{myer96,difr01,atta09}.
The schematic configuration of the model is illustrated in the Figure
4 of \cite{atta09}.

In the following discussion, the subscripts $f$, $r$, $B$, $R$ and $c$
are used to represent the front infall layer,
rear infall layer, blue-shifted outflow layer, red-shifted outflow
layer and the continuum core, respectively. On the assumption of
homogeneous and isothermal layers of gas, the line intensity or
brightness temperature from the ``four-layer complex'' can be solved
from the radiative transfer equation for given excitation temperatures
$T_{f}$ and $T_{r}$ of the front and rear layers of infall,
$T_{\rm out}$ of both the outflow layers, and $T_{c}$ of the
continuum emission. The line brightness temperature, after subtracting
the continuum in which the contribution from the cosmic background
radiation is negligible, can be expressed as:
\begin{eqnarray}
\Delta T_L
&=&J(T_{f}) \Bigg[ \left(1-\frac{J(T_c)f_c(1-e^{-\tau_c})}{J(T_{f})}\right)(1-e^{-\tau_f})+ \nonumber \\
&&\frac{J(T_r)}{J(T_f)}(1-e^{-\tau_r})e^{-\tau_f-\tau_c} \Bigg] e^{-\tau_B}+ \nonumber \\
&&J(T_{\rm out})\Bigg[ \left(1-\frac{J(T_c)f_c(1-e^{-\tau_c})}{J(T_{\rm out})}\right)(1-e^{-\tau_B})+ \nonumber \\
&&(1-e^{-\tau_R})e^{-\tau_B-\tau_f-\tau_c-\tau_r}\Bigg],
\end{eqnarray}
where ${\displaystyle
J(T)=\frac{h\nu}{k}\left(e^{h\nu/kT}-1\right)^{-1}}$ is a radiation
temperature of each layer and $f_{\rm c}$ is the beam filling factor
for the continuum emission. Equation 1 shows that, in the four-layer
model, the resultant line-profile can be expressed as two
separated terms. The term in the first  bracket on the right side of
Equation 1 describes the line profile of the in-falling gas,
corresponding to an inverse P-Cygni profile in the case of $J(T_{
c})f_c\tau_c$$>$$J(T_f)$. The infall line profile is
attenuated by the absorption in the blue-shifted outflow
($e^{-\tau_B}$). The term in the second bracket corresponds to the line
profile from the bipolar outflow showing a classical P-Cygni profile if
$J(T_{c})f_{c}\tau_c$$>$$J(T_{\rm out})$ or a red-dominated
double-peaked profile if $J(T_{c})f_{c}\tau_c$$<$$J(T_{\rm
out})f_{\rm out}$. The emission from the red-shifted outflow layer is
attenuated by the absorbing gas in both the infall layers and the
blue-shifted outflow as well as the continuum dust
($e^{-\tau_f-\tau_r-\tau_B-\tau_c}$).

Each of the optical depths in the four layers can be modeled using a
Gaussian function with a mean velocity and a velocity dispersion for
infall ($V_{\rm in}$, $\sigma_{\rm in}$) and outflow ($V_{\rm out}$,
$\sigma_{\rm out}$) assuming the systemic velocity of $V_{\rm sys}$.
Therefore, a total of fourteen parameters, namely, the five kinematic
parameters ($V_{\rm in}$, $V_{\rm out}$, $\sigma_{\rm in}$,
$\sigma_{\rm out}$ and $V_{\rm sys}$) along with four excitation
temperatures ($T_f$, $T_r$, $T_{\rm out}$ and $T_c$)
and five optical depths ($\tau_f$, $\tau_r$, $\tau_{
B}$, $\tau_R$ and $f_c$$\tau_c$) are required to
characterize the physical conditions of the gas in each layer.

\subsubsection{Fitting the HCN(3-2) Spectra}

We made a tentative fitting to the observed HCN(3-2) spectra with
the model described by Equation 1. In the case of W3-SE, the
quantities for the continuum emission ($T_c$ = 40 K and $\tau_c$
= 0.5) can be determined from the SED as discussed by \cite{zhu10}
assuming that the ratio of 2:1 in dust content between SMA-1 and
SMA-2 is proportional to the ratio of the total between the two
sources. Therefore, thirteen free parameters are needed.

In order to fit the four-layer model to the observed HCN(3-2) line
profiles, we first assumed the two outflow layers are optically
thin at low velocities and fit the low-velocity part (from
--43.5 to --34.1 km s$^{-1}$), or the assumed infall part, with the
least-square approach. The high-velocity parts of
the blue wings ($<$ --43.5 km s$^{-1}$) and red ($>$ --34.1
km~s$^{-1}$) of HCN(3-2) are related to the bow-shock, and could not
be fitted with Equation 1. Therefore, they are fitted separately with
a power-law profile (${\displaystyle
\propto~|V-V_{\rm sys}|^{-\gamma}}$). The maximum intensities of the
power-law line wings are set to be at --43.5 and --34.1 km s$^{-1}$
for blue and red wings, respectively, and the values are obtained
from fitting the low-velocity line profile with Equation 1.

The best fitting results are shown in Figure 11a, which gives an
infall velocity 0.9 km s$^{-1}$. The derived excitation temperatures
are 6$^{+1}_{-2}$ and 11$\pm3$ K for the front and rear layers,
respectively. The best fits for the thirteen free parameters in
the final stage was adjusted by eye to achieve a minimum residual.
The errors (1$\sigma$) were assessed by adding an offset (3$\sigma$)
to a best-fitted quantity with other parameters fixed, to increase
the offset of the best-fitted curve from the observed profile (or a
factor of 3 increase in the residual). The higher excitation
temperature for the rear layer might suggest the presence of a
temperature gradient in radius (temperature increases as it gets
close to the protostar). Thus, an observer receives emission from
colder part in the front and warmer part in rear layers,
considering the optical-depth effect. On the other hand, the
excitation temperature of the in-falling gas is also constrained
by the observed brightness temperature in the presence of an
inverse P-Cygni profile. As both angular and spectral resolution
improve, observations with interferometer arrays at millimeter and
sub-millimeter wavelengths will place more meaningful constraints
on the in-falling gas in proto-stellar cores. For the bipolar
outflow in SMA-1, the gas components in both the blue- and
red-shifted outflows are optically thin ($\tau_{\rm out}\sim$0.1)
with excitation temperatures of $\sim$75 K, which are much higher
than the beam-averaged peak brightness temperature ($\sim$8 K) of
HCN(3-2). As for the high-velocity parts of the spectrum,
the power-law index $\gamma$ for the blue and red lobes of
the outflow are 2.0 and 2.5, respectively. The difference of the two
values could be due to a slight difference in the directions of the
ejected blue- and red-shifted high-velocity gas components.

Fitting the four-layer model to the spectral profile of SMA-2
suggests that a bipolar outflow might also be present in this
proto-stellar core (see Figure 11b). In comparison, the expected
outflow from SMA-2 has a similar optical depth and excitation
temperature to those derived from SMA-1 but it appears at an
earlier stage. The observed line profile in the low-velocity
range appears to be dominated by outflow emission and no obvious
infall signatures are present toward SMA-2. SMA-2 is different
from SMA-1 since it has no obviously associated mid-IR point sources.
The relative extended mid-IR emission in the adjacent of SMA-2 might
be due to the external heating from the formed star in the W3-SE region.

We compared these fits with HCN(3-2) data with previous fits to
CARMA HCO$^+$(1-0) data of lower angular resolution
($\sim$6$\arcsec$) using a similar four-layer model \citep{zhu10}.
All the parameters from the HCN(3-2) fits (see the caption of
Figure 11) are consistent with the previous HCO$^+$(1-0) fits,
especially the infall velocity (0.9 km s$^{-1}$). Different from
the HCO$^+$(1-0) fits, the HCN(3-2) suggests an inverse P-Cygni
profile instead of a blue profile, which could be due to the
larger contribution from the high-velocity outflow in HCN(3-2)
and the significant improvement of angular resolution.

We found it is implausible to obtain consistent fits with Equation
1 for the entire W3-SE region, and the fitted values of $\tau_f$,
$\tau_r$, $T_f$ and $T_r$ change dramatically at difference
positions. It could partially due to the contaminations from
other components, such as potential further multiplicity, the
possible large-scale outflow, and the possible high-velocity
outflow associated with SMA-2. A more essential cause could be
the over-simplified assumptions of the model itself, which
assumes complete overlaps of the layers along the light of sight.
A more detailed model should consider the two dimensional
geometry of the W3-SE region and include the influences of the
core SMA-2, the additional HCN(3-2) blue-shifted component and
other possible components. Although due to the limit of the model
we can only qualitatively analyze most of the fitted physical
parameters, an exception is the infall velocity ($V_{\rm in}$).
It can be consistently fitted for the spectra from different
positions; its value is always between 0.8 and 0.9 km s$^{-1}$ and
it is not sensitive to other fitted parameters. Actually it is
also consistent with the velocity difference between the HCN(3-2)
emission ``bump" and the assumed systemic velocity.

Although fitting the HCN(3-2) spectrum towards SMA-1 is a
tentative exercise, the fitting demonstrates a possible explanation:
a complex including both high-velocity wings and an inverse
P-Cygini profile, for the observed HCN(3-2) line profile toward
SMA-1 region. A multiple-outflow model for the observed profile
toward SMA-1 can be ruled out since no absorptions against the
continuum are expected from the outflow gas with an excitation
temperature of $\sim75$ K, which is an order of magnitude larger
than the brightness temperature of the continuum emission ($8$ K)
at the observing wavelength. However, an alternative model with
multiple outflows together with a cold envelope might fit the
observed spectral profile. High-resolution observations of dense
molecular lines at higher frequencies and a more sophisticated
model are needed to unambiguously determine the nature of the
observed profile towards SMA-1.

\subsection{The Nature of the Double Cores in W3-SE}

The W3-SE dust core is associated with a total mass of 65$\pm$10
M$_\sun$ in a 15$\arcsec$  region, and is characterized by two
continuum emission sources SMA-1 and SMA-2 with a projected
angular separation 3.5$\arcsec$ or 7000~AU. SMA-1 and SMA-2
contribute 1/2 and 1/4 of the total flux density in 1.1-mm
continuum, respectively \citep{zhu10}. Assuming a uniform
distribution for both dust temperature and
dust-to-gas ratio over the entire W3-SE region, a mass of 32$\pm$5
M$_\sun$ is inferred for SMA-1 and 16$\pm$2 M$_\sun$ for SMA-2.
In the vicinity of the double proto-stellar cores, at least three
additional mid-infrared sources have been revealed by {\it
Spitzer}, indicating that the double cores located within or near
SMA-1 and SMA-2 might be a young cluster \citep{zhu10}. Both the
SMA and CARMA observations of the molecular lines reveal a common
filamentary envelope of molecular gas surrounding the two main
proto-stellar cores, suggesting that SMA-1 and SMA-2 result from
fragmentation of molecular gas in the same parent cloud.

The dust temperature of the W3-SE core region (covering SMA-1 and
SMA-2) $\sim$40 K, indicating that the gas in the core region is
likely to be heated by both internal and external energy sources.
Indeed, a group of mid-IR sources are located east of SMA-1 and
SMA-2, and mid-IR emission is also detected in SMA-1
\citep{zhu10}. The high temperature implies a relatively large
Jeans mass. Assuming a mass of 65 M$_\sun$ evenly distributed in
a  15$\arcsec$ region ($\sim$0.15 pc), a Jeans mass could be
estimated by
\begin{displaymath}
M_{\rm J} = 1.2\times10^5(\frac{T}{100~K})^{3/2}(\frac{\rho}{10^{-24}~{\rm g}~{\rm cm}^{-3}})^{-1/2}{\mu}^{-3/2}~M_\sun
\end{displaymath} \citep{lang98}, where $\mu$ is the mean
molecular weight and assumed to be 2.35.
The inferred Jeans mass is $\sim$7 M$_\sun$, far less than the derived
masses of SMA-1 and SMA-2.

The observed infall signature might be due to global infall and/or
inflow toward SMA-1. Although the spectra of HCN(3-2) and
HCO$^+$(3-2) appear to be dominated by outflow, they still present
significant red-shifted absorption (see Figure 7). The HCO$^+$(1-0)
spectrum toward the W3-SE core, which covers both SMA-1 and SMA-2,
also shows the infall signature \citep{zhu10}. A global infall in
W3-SE is likely to occur as the inferred Jeans mass is far less
than the total mass of the W3-SE molecular core.
However, both the spectral map (Figure 3) and the integrated
intensity map (Figure 4b) show that the possible inverse P-Cygni
profile in HCN(3-2) is present toward SMA-1, the major one of the
double cores in W3-SE.

SMA-1 and SMA-2 may result from fragmentation of a common parent
molecular cloud. The mechanisms for binary formation have been
discussed by \cite{tohl02}. The location of SMA-1 and SMA-2
within a common filament (Figure 2c) suggests
that the two cores result from a prompt
fragmentation mechanism \citep{tohl02}. Considering the projected
separation, $d$ = 7000 AU (3.5$\arcsec$) for SMA-1 and SMA-2, and
their total mass, $M_{\rm tot}\approx50$ M$_\sun$, they could be
gravitationally bound if their actual separation is comparable to
projected one. Assuming SMA-1 and SMA-2 rotate around their mass
center with a distance $d$, the orbital period
{$\displaystyle P = (2\pi {d^3})^{1/2} \left(GM_{\rm tot}\right)^{-1/2}\approx3\times10^4$}
yr and the relative velocity of the two cores $\sim$2.5
km~s$^{-1}$. Although no significant difference in the radial
velocities of the double cores is seen in our observations, this
could be explained if the orbit is nearly face-on as implied by
the large inclination angle ($\sim$70$\arcdeg$) of the
high-velocity outflow in HCN(3-2) originating from SMA-1
(Section 4.1). The W3-SE proto-stellar system appears to be very
young with an age $\sim3\times$10$^3$ yr inferred from the
kinematic timescale of the outflow. This is an order magnitude
smaller than the orbital period if the molecular core W3-SE is
to form a protobinary system, indicating it is still an
unrelaxed system. However, the possibility that SMA-1 and SMA-2
will relax to a gravitationally-bound binary system cannot be
ruled out.
Accretion could  decrease the separation of the double cores and help
the formation of a binary as suggested in the theory of dynamical hardening \citep{bonn05}.

W3-SE can be compared with  NGC 1333 IRAS 4, which was reported
as a binary system consisting of two proto-stellar cores 4A
(9.2 M$_\sun$) and 4B (4.0 M$_\sun$) with an angular separation
$\sim$30$\arcsec$ ($\sim$9000 AU) at a distance of $\sim$300 pc
\citep{sand91}. In follow-up observations with higher-angular
resolution, further multiplicities in both the sub-cores 4A and
4B have been  revealed \citep[e.g.][]{lay95,loon00}. Given the
large masses for both SMA-1 and SMA-2 in W3-SE and the limited
angular resolutions of the existing studies, SMA-1 and SMA-2
might turn out to be composed of multiple sub-cores in
higher-resolution observations. The observed adjacent IR sources
\citep{zhu10} and multiple outflows also suggest further
multiplicity in SMA-1 and SMA-2.

\section{Summary}

Following our previous study with continuum and spectral lines at
low angular resolution, we made high-resolution observations of
multiple molecular lines toward the double proto-stellar cores
SMA-1 and SMA-2 in W3-SE. The two proto-stellar cores are located
in a molecular filamentary structure which shows an active
star-forming region in W3-SE. A high-velocity bipolar outflow
originating from SMA-1 is seen in HCN(3-2), with confirmation
from other molecular lines, which shows a star formation activity
in W3-SE. The morphology and the velocity
profile of the bipolar outflow fit a bow-shock outflow model.
SMA-1 and SMA-2 appear to result from fragmentation of a
collapsing massive molecular core. Global collapse and gas infall
are suggested by red-shifted self-absorption observed in both
HCN(3-2) and HCO$^+$(3-2). An inverse P-Cygni feature observed in
the HCN(3-2) line toward the proto-stellar core SMA-1 suggests
that SMA-1 is the major accretion source in W3-SE. The double
proto-stellar cores in W3-SE are ideal targets for further
observations with higher angular and spectral resolution at
(sub-)millimeter wavelengths in order to understand the formation
of clusters with intermediate-mass stars.

\acknowledgments

We are grateful to the staff of the SMA and CARMA telescopes who
make these observations possible. We also thank the anonymous
referee for his detailed comments and questions which have
significantly improved the presentation of these results. We thank
Wu Y. for the suggestions and discussions in forming the idea of
this paper, selecting the research targets and providing the
single-dish data of the targets, suggesting the suitable
molecular probes, as well as the discussions in the analysis of
the results. L. Z. is supported by the SAO pre-doctoral program
during this research, and he is also supported by Grant 10873019
at NSFC when staying at Peking University.

\clearpage

\appendix
\section{Bow-shock Model for Outflow}

The bow-shock model \citep{down99} is one of the two popular models
that provide mechanisms to accelerate molecular gas in an outflow. The
observed features of the high-velocity bipolar outflow from SMA-1
appears to show the characteristics of a bow-shocked outflow as
described in both numerical simulations and analytical models. In order
to fit the P-V diagram from the observed HCN(3-2) outflow with the
bow-shock model, we have worked out the formulas to outline the
boundary of a P-V diagram from a bow-shock surface based on the
analytical model of \cite{down99}. In this model the bow-shock
discontinuity is described as a rigid surface moving in the
interstellar medium with a value of bulk velocity $v$. For instance, in
a cylindrical coordinate system, the locus of the bow shock in a
blue-shifted outflow lobe can be described as:
\begin{equation}
z = a-\mid{r}\mid^s, ~s{\ge}2,
\end{equation}
where $z$ is an axis along the direction of the bulk velocity of the
bow-shock surface, $r$ is radius perpendicular to axis $z$, and $s$ is
a power-law index to describe the curvature of the bow-shock surface
(see Figure 10). In addition, the bow-shocked surface is treated as a
rigid body. We made the further assumptions used in the following
reduction. In the bow-shock rest frame, ambient medium moves towards
the bow-shock with a value of velocity $v$. As soon as the ambient
molecular gas hits the surface of the bow-shock, the motion of the
ambient gas alters the direction that is tangential to the bow-shock
surface with a value of velocity $v_1$:
\begin{equation}
v_1 = \frac{v}{\sqrt{1+R^2}}\sqrt{\frac{1}{16}+R^2},
\end{equation}
where ${\displaystyle R\equiv\Big| \frac{v_z}{v_r}
\Big|=\Big|\frac{{\rm d}z}{{\rm d}r}\Bigr|=s(a-z)^{(s-1)/s}}$, $v_z$
and $v_r$ are the components of the velocity $v$ projected on the $z$
and $r$ directions (in the bow-shock fixed frame), and $a$ is the
projected distance between bow-shocked tip and the origin of the
outflow.

Considering the Doppler shift and the geometry described above, a
complete surface of a bow-shocked bipolar outflow lobes can be divided
into four separated parts, namely a combination of front and rear
layers for each of the blue- and red-shifted lobes. Each of the four
boundaries corresponds to a curve in a quadrant of the P-V space to
outline a P-V diagram. Therefore, four separate  functions $v'_{Bf}$,
$v'_{Rf}$, $v'_{Br}$ and $v'_{Rr}$ are used to describe the boundaries
of bipolar outflow in the P-V diagram. Hereafter, the subscripts $B$
or $R$ represent for blue or red velocities and $f$ or $r$ for front
or rear layers of a given bipolar outflow.

Giving a front layer of a blue-shifted bow shock, the observed radial
velocity $v^\prime_{Bf}$ in the LSR frame at a point of the bow shock
surface is a function the tangential velocity $v_1$:
\begin{eqnarray}
v^\prime_{\rm Bf} =
\frac{v_1}{\sqrt{1+R_B^2}}(-\cos\alpha+R_B\sin\alpha)-v\sin\alpha+V_{\rm
sys},
\end{eqnarray}
where $V_{\rm sys}$ is the systemic velocity of the source originating
the bow-shock outflow, and $\alpha$ is the inclination angle, the angle
between the axis of the outflow and the plane of sky.

Therefore, substituting $v_1$ with Equation A2, the observed radial
velocity at a given point (Equation A3) can be re-written as:
\begin{eqnarray}
v'_{\rm Bf} =
\frac{v}{1+R_B^2}(-\cos\alpha+R_B\sin\alpha)\sqrt{\frac{1}{16}+R_B^2}-v\sin\alpha+V_{\rm sys},
\nonumber \\
v'_{\rm Rf} =
\frac{v}{1+R_R^2}(-\cos\alpha-R_R\sin\alpha)\sqrt{\frac{1}{16}+R_R^2}+v\sin\alpha+V_{\rm sys},
\nonumber \\
v'_{\rm Br} =
\frac{v}{1+R_B^2}(\cos\alpha+R_B\sin\alpha)\sqrt{\frac{1}{16}+R_B^2}-v\sin\alpha+V_{\rm sys},
\nonumber \\
v'_{\rm Rr} =
\frac{v}{1+R_R^2}(\cos\alpha-R_R\sin\alpha)\sqrt{\frac{1}{16}+R_R^2}+v\sin\alpha+V_{\rm sys},
\end{eqnarray}
where ${\displaystyle R_B=s_B(a_B-z)^{(s_B-1)/s_B}}$ and
${\displaystyle R_R=s_R(a_R+z)^{(s_R-1)/s_R}}$. The values of the
bow-shock velocities for the blue- and red-shifted outflow lobes have
been assumed to be the same in Equation A4. This assumption is valid
since the line widths of the blue- and red-shifted wings are very
similar as discussed in Section 3.1. We can further simplify the model
for the velocity profiles by assuming $s_R = s_B = 2$. In addition, the
projected distances between the bow shock tips and the origin of the
outflow ($a_B$ and $a_R$) can be directly determined from the outflow
image.  The systemic velocity $V_{\rm sys}$ can be determined from the
peak velocity of hot molecular lines. For SMA-1, $V_{\rm sys}$ = --39.1 km
s$^{-1}$ is used  (see Section 3.4). Thus, the remaining free
parameters in Equation A4 are the inclination angle $\alpha$ and the
bow-shock velocity $v$.

Equation A4 describes the observed radial velocities along the
boundaries of a bow-shock outflow as a function of $z$. Given a
cylindrical coordinate system with an origin that is the same as
that of the outflow (as illustrated in Figure 10), the position
on the sky plane ($r'$, $z'$) can be transferred from the
coordinates ($r$, $z$) on the bow-shock surface for a given
inclination angle $\alpha$:

\begin{eqnarray}
z' = z\cos\alpha+r\sin\alpha
\nonumber \\
r' = -z\sin\alpha+r\cos\alpha
\end{eqnarray}

Using Equations A1 and A5, the sky position $z'$ can be related to
the outflow coordinate $z$ for each of the four boundaries:
\begin{eqnarray}
z'_{\rm Bf} = z\cos\alpha-(a_B-z)^{1/2}\sin\alpha
\nonumber \\
z'_{\rm Rf} = z\cos\alpha-(a_R+z)^{1/2}\sin\alpha
\nonumber \\
z'_{\rm Br} = z\cos\alpha+(a_B-z)^{1/2}\sin\alpha
\nonumber \\
z'_{\rm Rr} = z\cos\alpha+(a_R+z)^{1/2}\sin\alpha
\end{eqnarray}

Thus, the observed radial velocities (Equation A4) for a given
bow-shocked bipolar outflow can be implicitly related to the coordinate
$z'$ on the sky plane with the Equation A6 by eliminating the argument
$z$. With the four implicit functions $v'=v'(z')$, we can outline a P-V
diagram along the projected major axis of a bipolar outflow using the
bow-shocked model. Note that in the above deduction we worked out the
geometrical relation without including optical depth effect.

\clearpage
\begin{figure}[ht]
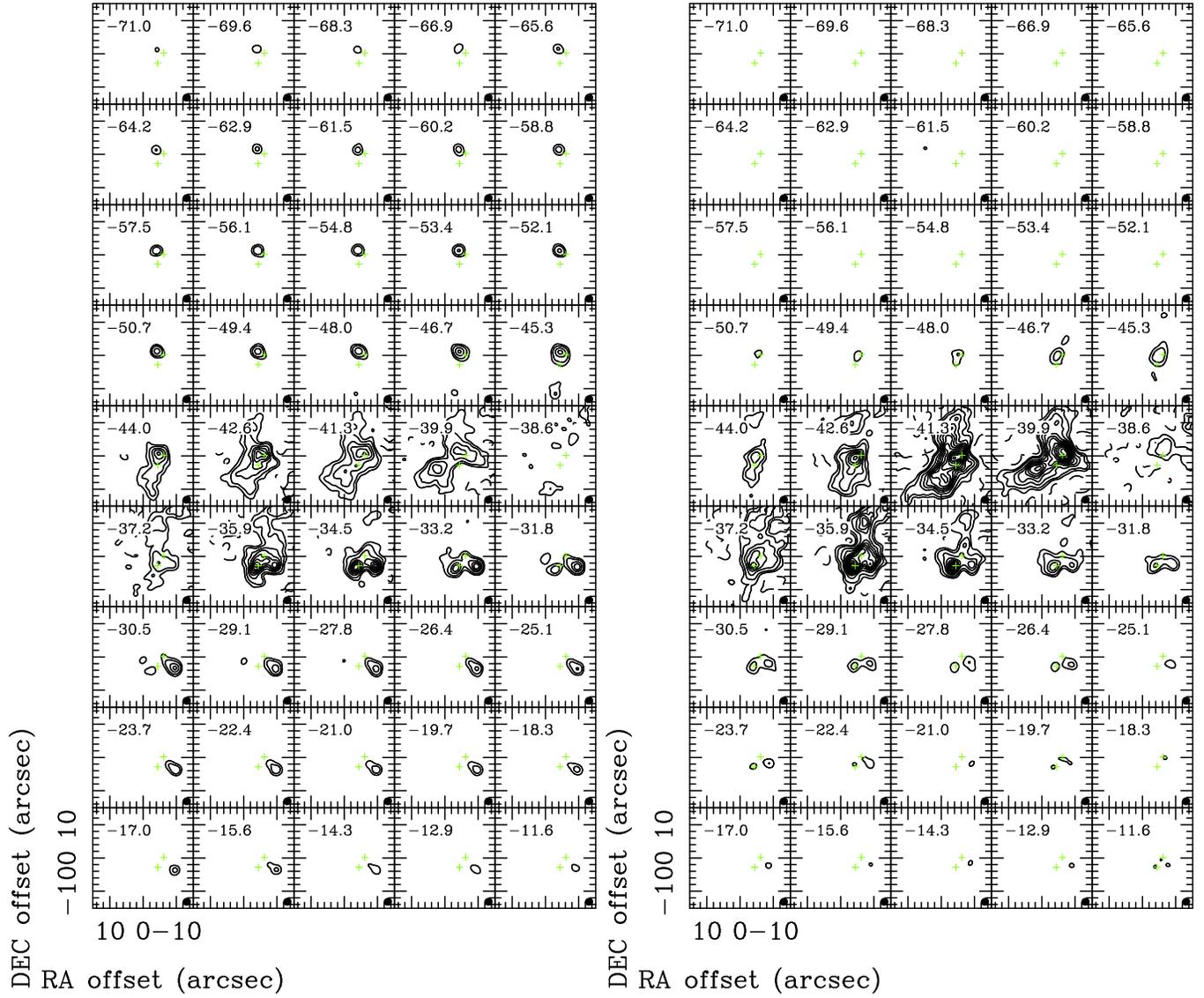

  \centering
  \includegraphics[width=15.0cm, angle=-90]{f1a_hcn.l.p.ps}
  \includegraphics[width=15.0cm, angle=-90]{f1b_hco.l.p.ps}
  \caption{The channel maps of HCN(3-2) (a: left panel) and HCO$^+$(3-2) (b: right panel)
  with a velocity interval of 1.4 km s$^{-1}$. Contours are (-1, 1, 2, 4,
  6,...)$\times5\sigma$, $\sigma$ = 0.045 Jy beam$^{-1}$ in both maps with an FWHM
  beam of 2.5$\arcsec$$\times$2.3$\arcsec$ (--65$\arcdeg$) as marked at right-bottom. The number at left-top in each channel image corresponds to the radial velocity. The green crosses mark the positions of the two proto-stellar objects, SMA-1 and
  SMA-2.}
\end{figure}

\begin{figure}[ht]
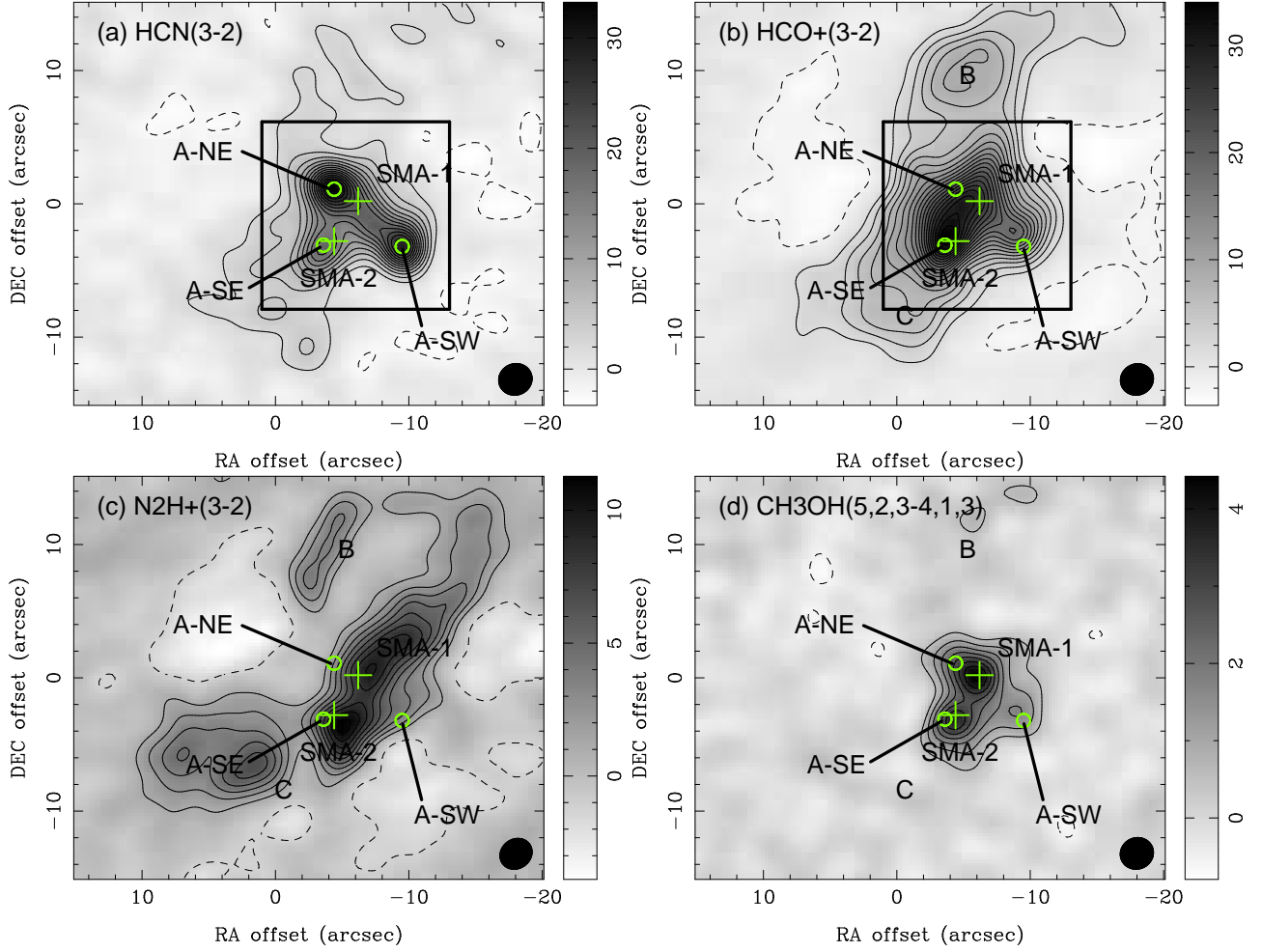

  \centering
  \includegraphics[width=6.5cm, angle=-90]{f2a_hcn.m0.all.p.ps}
  \includegraphics[width=6.5cm, angle=-90]{f2b_hco.m0.all.p.ps}
  \\
  \includegraphics[width=6.5cm, angle=-90]{f2c_n2h.m0.all.p.ps}
  \includegraphics[width=6.5cm, angle=-90]{f2d_ch3oh.m0.all.p.ps}
  \caption{Integrated intensity maps of (a) HCN(3-2) in the velocity range of
  (--70.5, --9.8) km s$^{-1}$, (b) HCO$^+$(3-2) in (--53.0, --21.0) km s$^{-1}$,
  (c) N$_2$H$^+$(3-2) in (--43.5, --34.5) km s$^{-1}$ and (d)
  CH$_3$OH(5$_{2,3}$-4$_{1,3}$) in (--44.5, --33.7) km s$^{-1}$. Contours are
  (--1,1,2,3,4...)$\times$3$\sigma$ Jy beam$^{-1}$ km s$^{-1}$. The rms noises (1
  $\sigma$) are $\sim$0.7, $\sim$0.7, $\sim$0.5 and $\sim$0.2 Jy Beam$^{-1}$ km
  s$^{-1}$ for the images (a), (b), (c) and (d), respectively. The images (a), (b)
  and (d) were made from the LSB data observed at $\nu_{\rm LO}=271.7$ GHz with an
  FWHM beam of $2.5\arcsec\times2.3\arcsec$ (--65$\arcdeg$). The FWHM beam of the
  image (c) (USB data) is $2.5\arcsec\times2.2\arcsec$ (--55$\arcdeg$). Crosses
  denote the positions of the proto-stellar objects SMA-1 and SMA-2. Open circles
  mark the three peak positions of the HCN(3-2) line emission (A-NE, A-SW and A-SE).
  The positions of the weaker emission components B and C observed in HCO$^+$(1-0)
  with the CARMA are also indicated.}
\end{figure}

\begin{figure}[ht]
  \centering
  \includegraphics[width=16.0cm, angle=0]{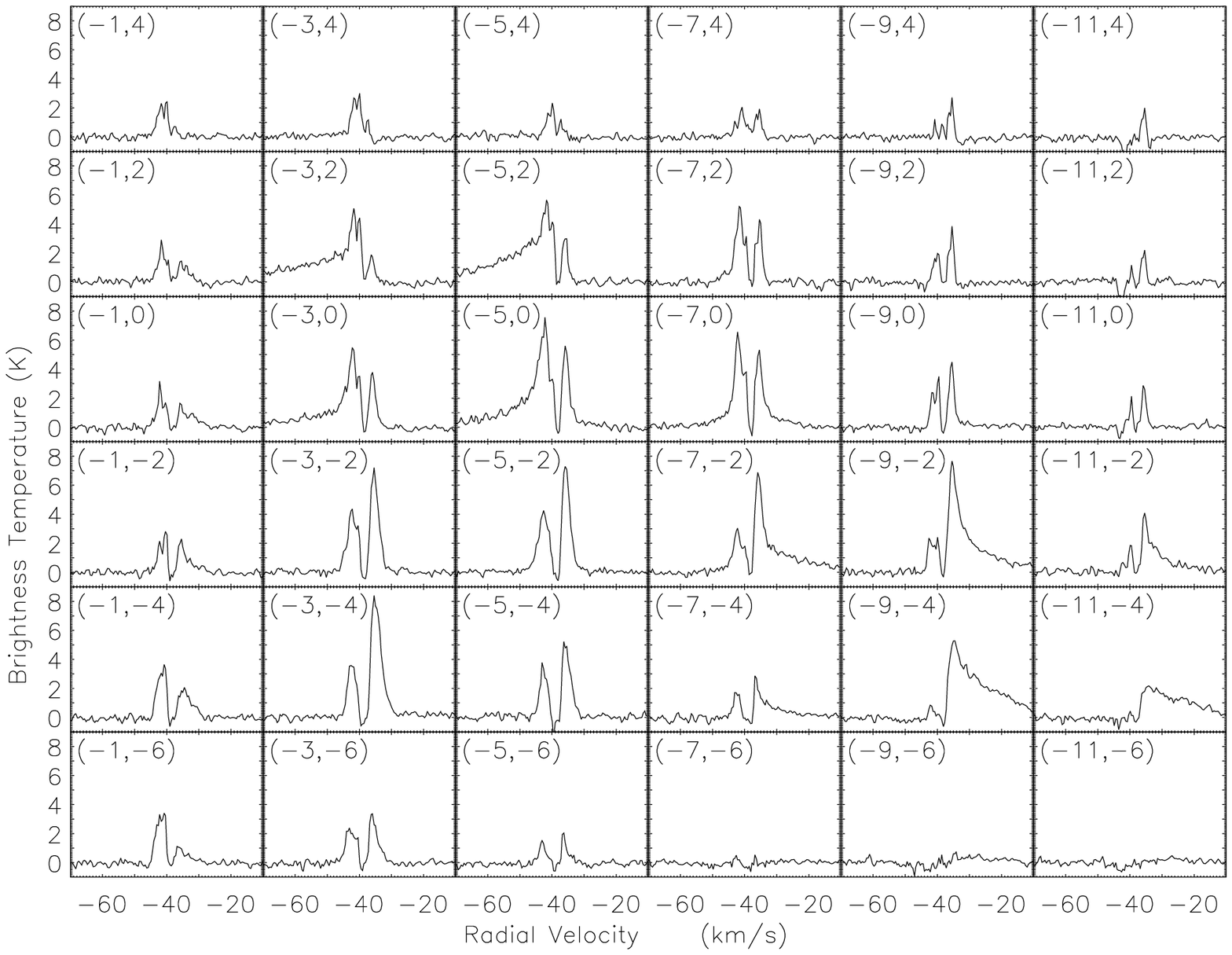}
  \caption{A grid of spectra for the HCN(3-2) line emission. The covered region is
  marked as the black rectangle in Figure 2a. The pair of numbers in the parenthesis
  at top-left in each of the spectra gives the offset in units of arcsecond with
  respect to the reference center.}
\end{figure}

\begin{figure}[h]
  \centering
  \includegraphics[width=6.0cm, angle=-90]{f4a_hcn.m0.wings.p.ps}
  \includegraphics[width=6.0cm, angle=-90]{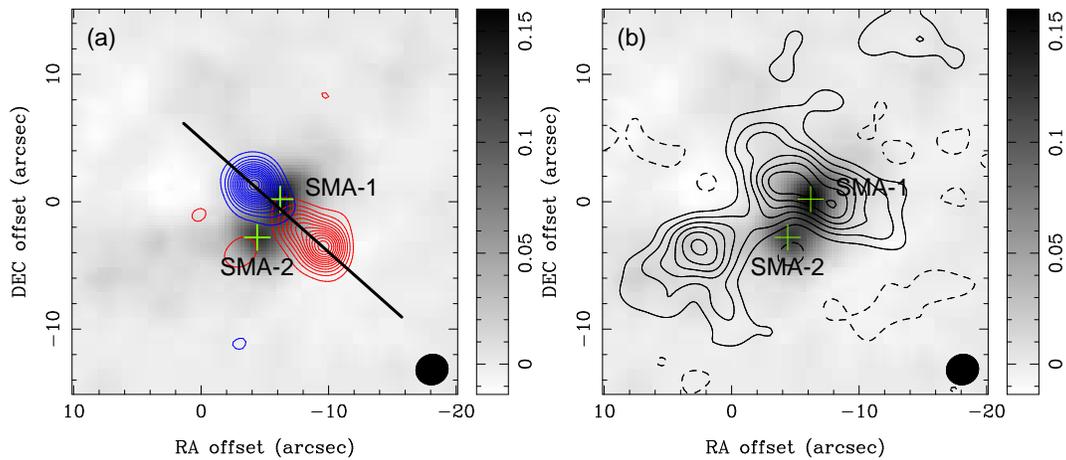}
  \caption{(a) The high-velocity outflow map in HCN(3-2), integrated from blue wing
  in the velocity range (--74.1, --45.2) km s$^{-1}$ (blue contours) and red wing in
  (--31.9, --5.8) km s$^{-1}$ (red contours). Contours are
  (-1,1,2,3,4...)$\times$5$\sigma$, 1 $\sigma$ = 0.28 Jy Beam$^{-1}$ km s$^{-1}$.
  The position angle of the outflow axis (the thick straight line) is 48$\arcdeg$.
  (b) The integrated intensity map of the blue-shifted HCN(3-2) emission bump in
  the velocity range of (--39.9, --39.5) km s$^{-1}$, which could somehow trace
  the distribution of possible infalling gas. Contours are
  (-1,3,2,3,4...)$\times$3$\sigma$, 1 $\sigma$ = 0.1 Jy Beam$^{-1}$ km s$^{-1}$.
  For both the figures the FWHM beam is 2.5$\arcsec$$\times$2.3$\arcsec$, P.A. =
  --65$\arcdeg$. For both figures the background shows the 1.1-mm continuum
  distribution, and the wedges are in unit of Jy beam$^{-1}$.}
\end{figure}

\begin{figure}[ht]
  \centering
  \includegraphics[width=16.0cm, angle=0]{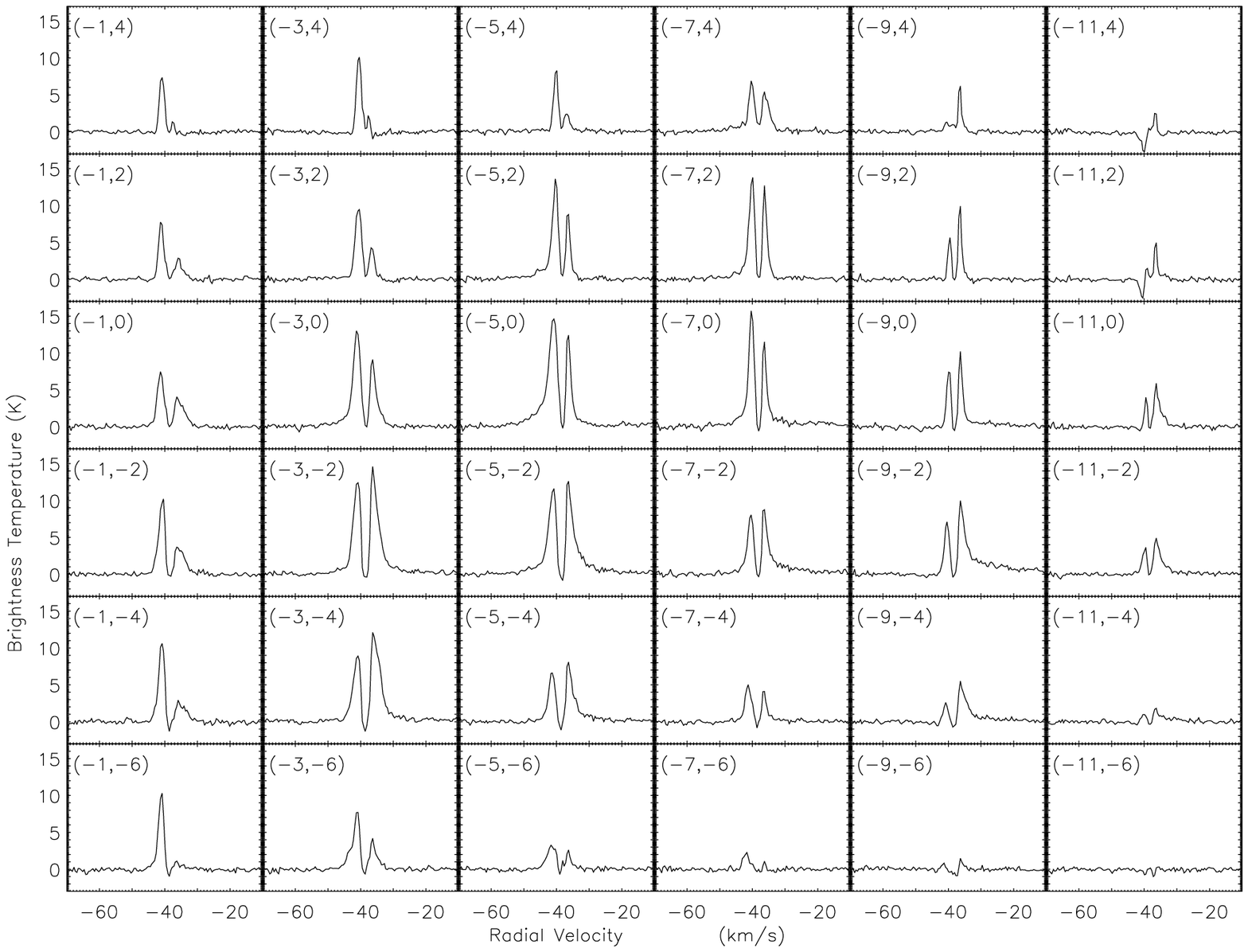}
  \caption{A grid of spectra for the HCO$^+$(3-2) line emission. The covered region
  is marked as the black rectangle in Figure 2b. The pair of numbers in the
  parenthesis at top-left in each of the spectra gives the offset in units of
  arcsecond with respect to the reference center.}
\end{figure}

\begin{figure}[h]
  \centering
  \includegraphics[width=6.0cm, angle=-90]{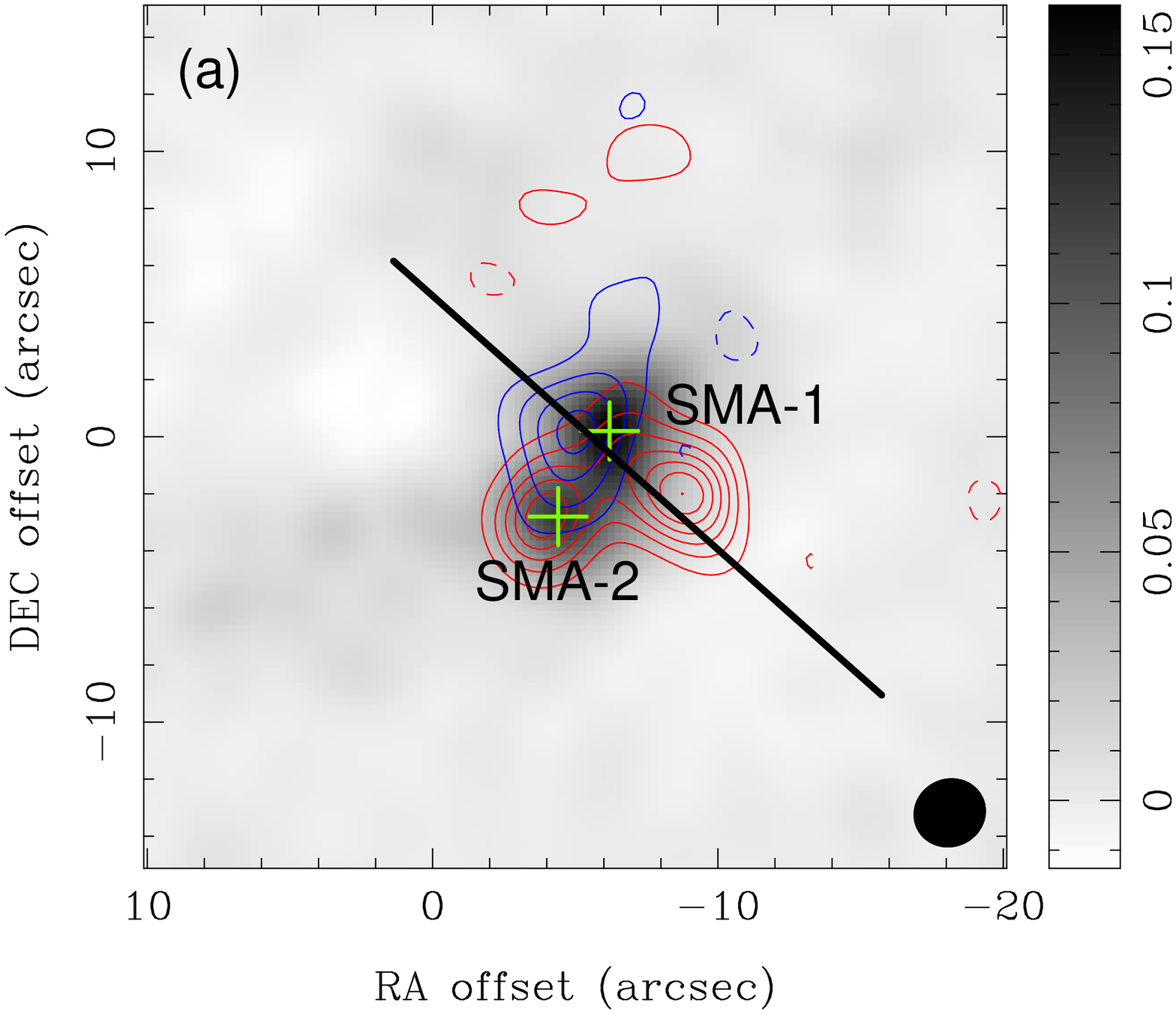}
  \includegraphics[width=6.0cm, angle=-90]{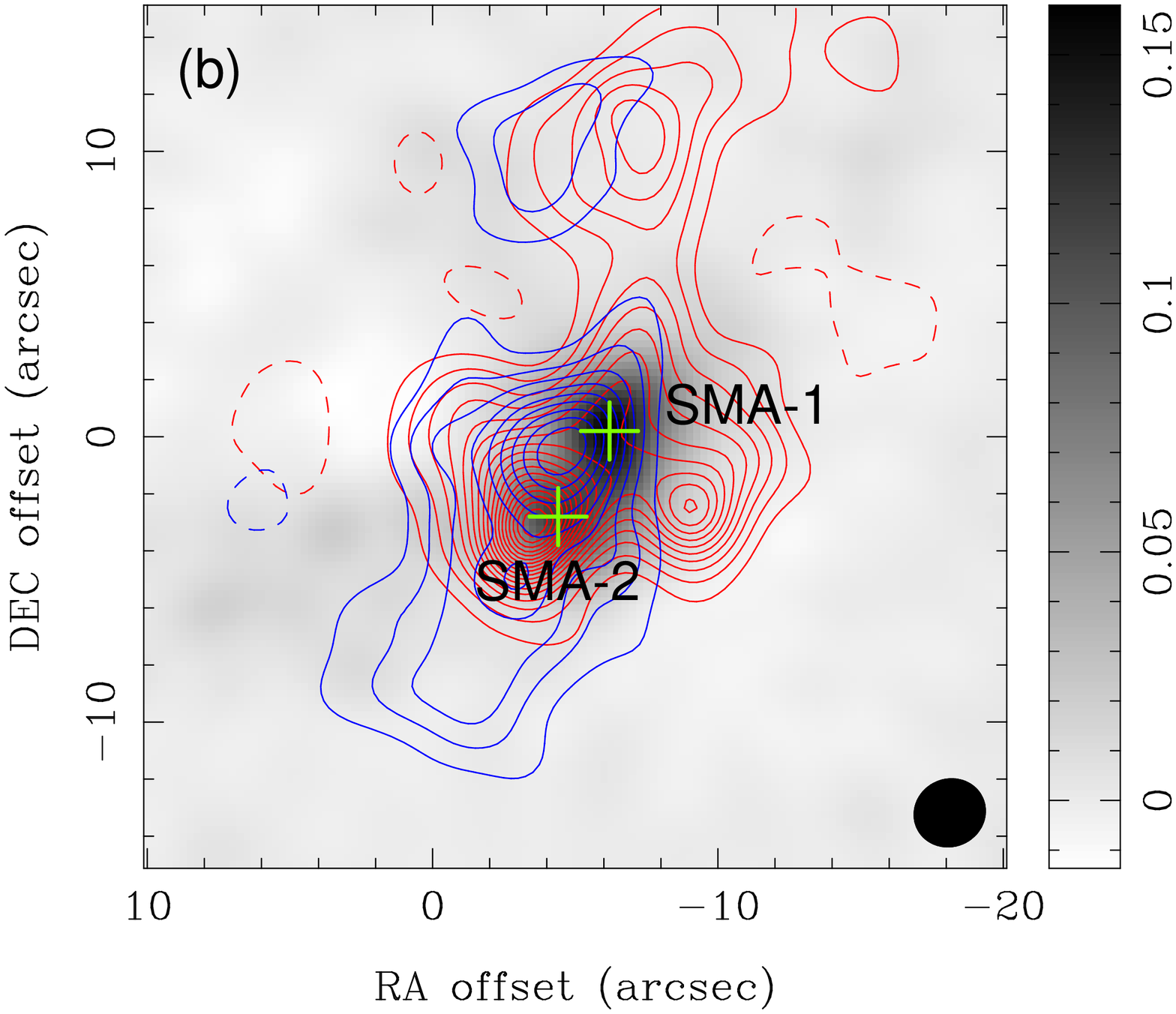}
  \caption{(a) The high-velocity outflow map in HCO$^+$(3-2), integrated from blue
  wing in the velocity range (--57.9, --45.3) km s$^{-1}$ (blue contours) and red
  wing in (--31.8, --18.8) km s$^{-1}$ (red contours). Contours are
  (-1,1,2,3,4...)$\times$4$\sigma$, 1$\sigma$ = 0.2 Jy Beam$^{-1}$ km s$^{-1}$.
  The thick straight line is the same as in Figure 4a. (b) The low-velocity outflow
  map in HCO$^+$(3-2), integrated from blue wing in (--44.9, --41.7) km s$^{-1}$
  (blue contours) and red wing in (--36.3, --32.3) km s$^{-1}$ (red contours).
  Contours are (-1,1,2,3,4...)$\times$5$\sigma$, 1$\sigma$ = 0.16 Jy Beam$^{-1}$ km
  s$^{-1}$. For both the figures the FWHM beam is 2.5$\arcsec$$\times$2.3$\arcsec$,
  P.A. = --65$\arcdeg$. For both figures the background is the 1.1-mm continuum image,
  and the wedges are in unit of Jy beam$^{-1}$.}
\end{figure}

\begin{figure}[h]
  \centering
  \includegraphics[width=12.0cm]{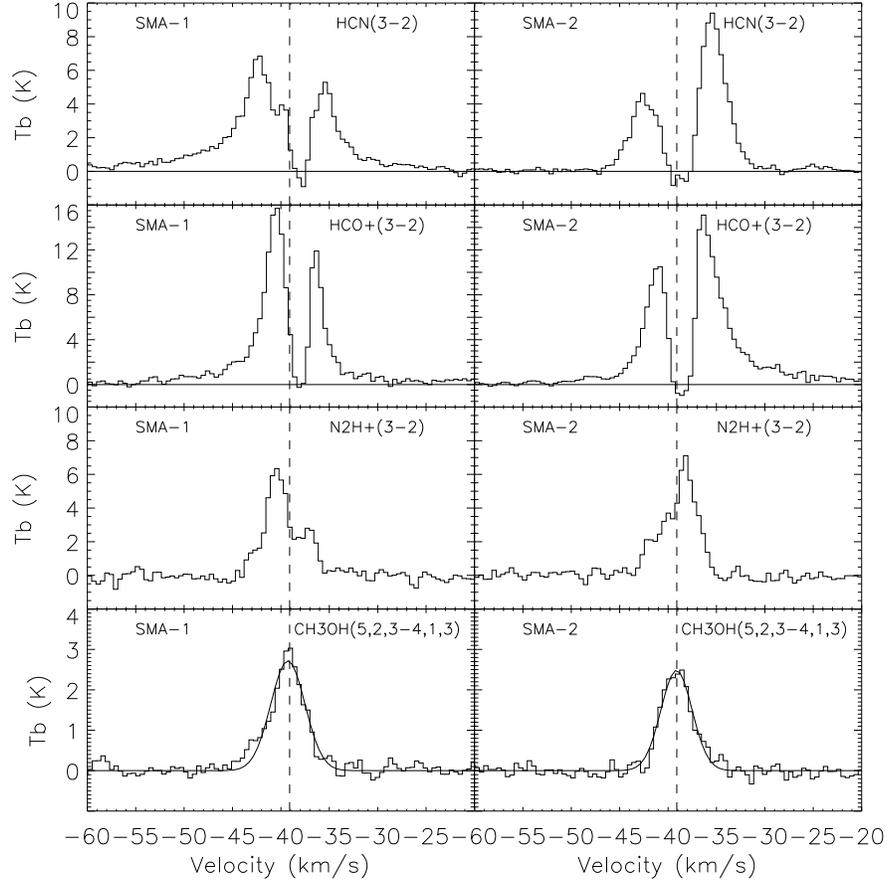}
  \caption{The spectra of HCN(3-2), HCO$^+$(3-2), N$_2$H$^+$(3-2) and
  CH$_3$OH(5$_{2,3}$-4$_{1,3}$) (from top to bottom) at SMA-1 (left panels) and SMA-2 (right panels). The vertical dash line in the two figures denotes the peak
  velocities of the CH$_3$OH(5$_{2,3}$-4$_{1,3}$) lines (--39.1 km s$^{-1}$), which
  is used as the systemic velocity through the paper.}
\end{figure}

\begin{figure}[h]
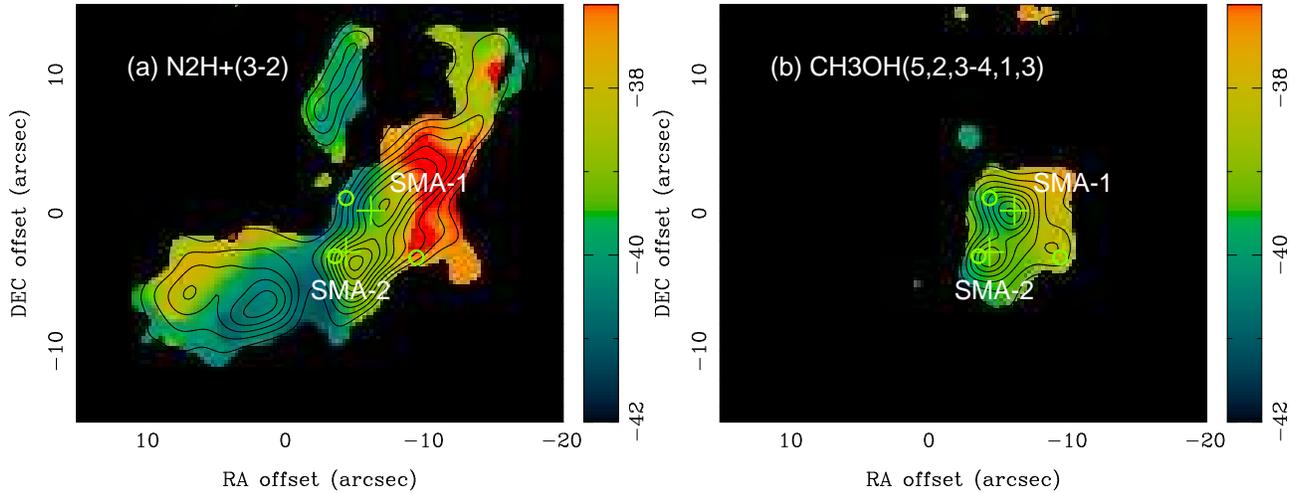

  \centering
  \includegraphics[width=6.5cm, angle=-90, origin=c]{f8a_n2h.m1.all.ps}
  \includegraphics[width=6.5cm, angle=-90, origin=c]{f8b_ch3oh.m1.all.ps}
  \caption{The centroid velocity maps of (a) N$_2$H$^+$(3-2) and (b)
  CH$_3$OH(5$_{2,3}$-4$_{1,3}$). The contours represent the integrated intensity of
  the line emission with values of (--1,1,2,3,4...)$\times3\sigma$ Jy beam$^{-1}$ km
  s$^{-1}$ for the given $\sigma=0.5$ and 0.2 Jy beam$^{-1}$ km s$^{-1}$ with an FWHM beam of $2.5\arcsec\times2.3\arcsec$ (--65$\arcdeg$) and
  $2.5\arcsec\times2.2\arcsec$ (--55$\arcdeg$) in images (a) and (b), respectively.
  The color wedge scales the radial velocity in the unit of km s$^{-1}$.}
\end{figure}

\begin{figure}[h]
  \centering
  \includegraphics[width=17.0cm, angle=0, origin=c]{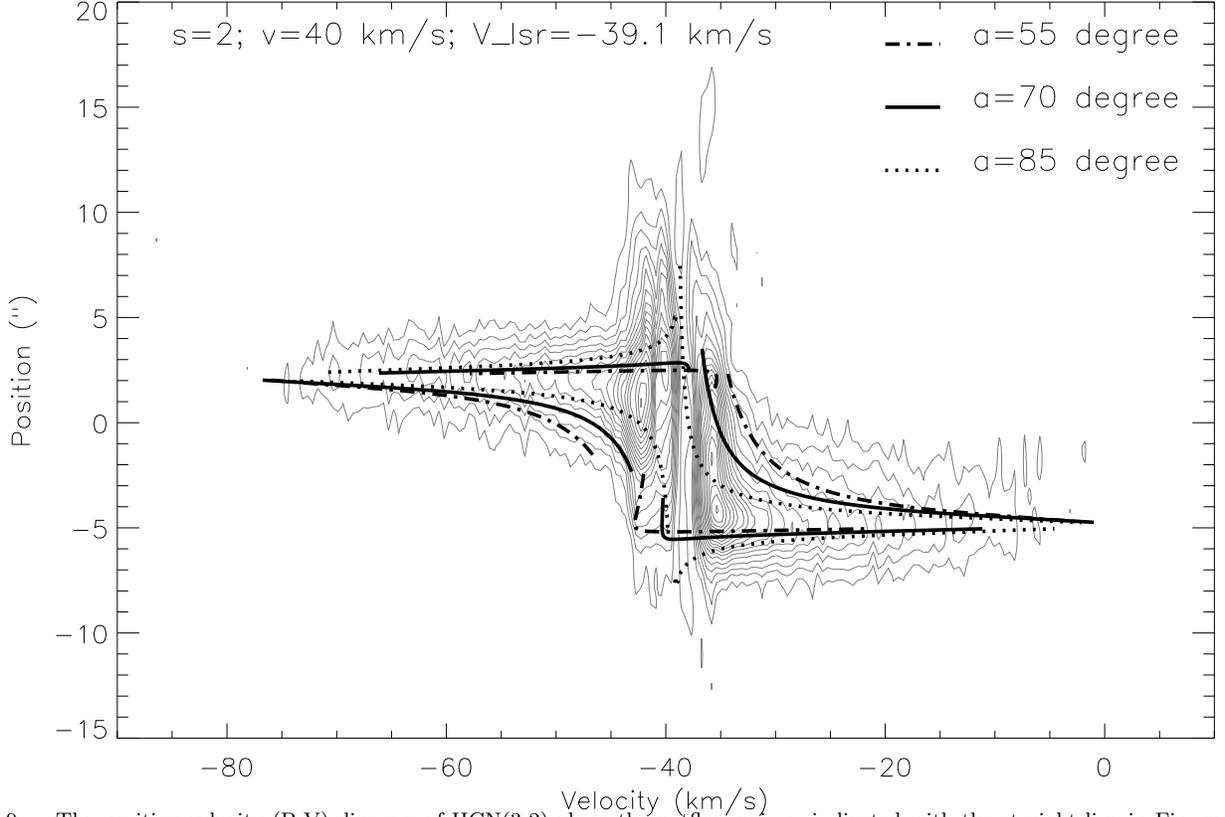}
  \caption{The position-velocity (P-V) diagram of HCN(3-2) along
  the outflow axis as indicated with the straight line in Figure
  4. The origin in the position axis corresponds to the position
  of SMA-1 projected on the axis. Contours are 0.2, 0.4, 0.6, 0.8,
  ... Jy beam$^{-1}$. The angular and velocity resolutions are
  2.7$\arcsec$ and 0.45 km s$^{-1}$, respectively. The observed
  P-V diagram is fitted with the bow-shocked outflow model as
  discussed in Appendix assuming the systemic velocity in the LSR
  frame ($V_{\rm sys}=-39.1$ km s$^{-1}$), the velocity of the
  bow-shock ($v=40$ km s$^{-1}$), and the spectral index of the
  power law for the bow-shock surface as function of radius
  ($s$ = 2). The dashed-dot-dashed, solid, dashed lines stand for
  the model curves corresponding to the three different values of
  inclination angle ($\alpha$=55$\arcdeg$, 70$\arcdeg$ and
  85$\arcdeg$). The curve with $\alpha$=70$\arcdeg$ appears to
  represent the best fitting. Note that bow-shock outflow model
  shows a good fit to the high-velocity gas. The low-velocity gas
  is probably disturbed by additional processes other than
  outflow. For example, the emission in the top-left quadrant in
  the velocities of (--40, --45) km s$^{-1}$ and the position
  offsets of (5$\arcsec$, 6$\arcsec$) are likely from the infalling
  gas or the gas in the molecular envelope, which is not related
  to the bow-shock outflow.}
\end{figure}

\begin{figure}[ht]
  \centering
  \includegraphics[width=10.0cm, angle=0]{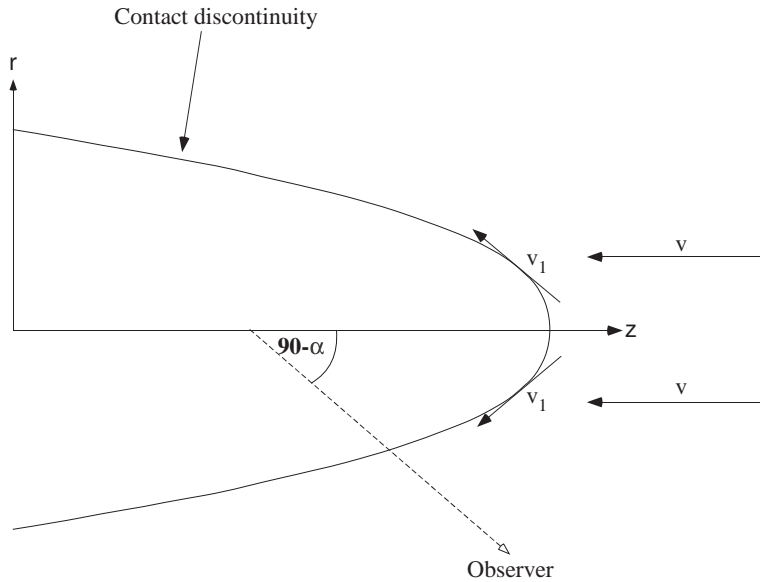}
  \caption{The sketch illustrates the geometry of the bow-shock
  outflow model, which is taken from the Figure 6 in \cite{down99}.}
\end{figure}

\begin{figure}[ht]
  \centering
  \includegraphics[width=15.0cm, angle=0]{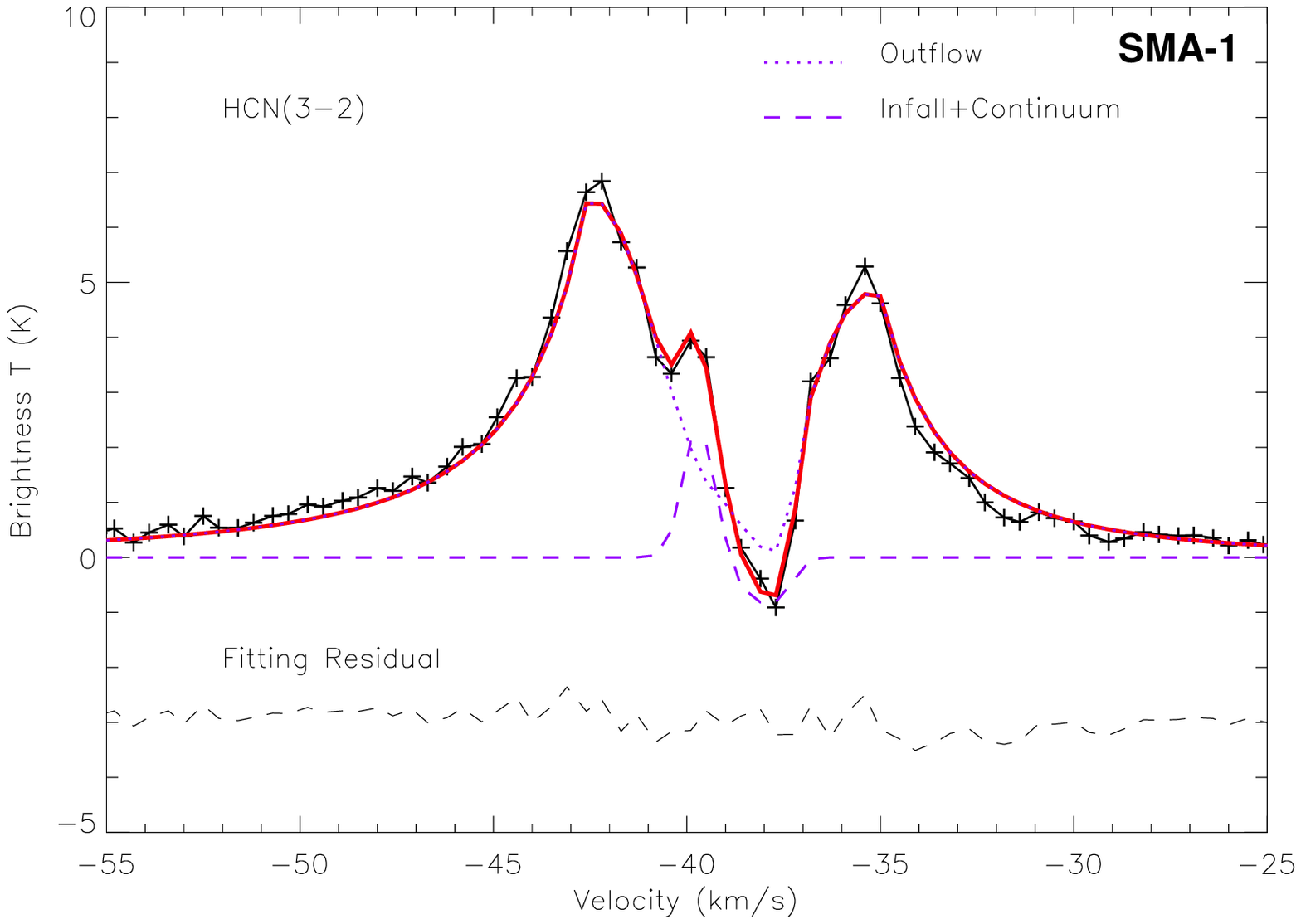}
  \includegraphics[width=15.0cm, angle=0]{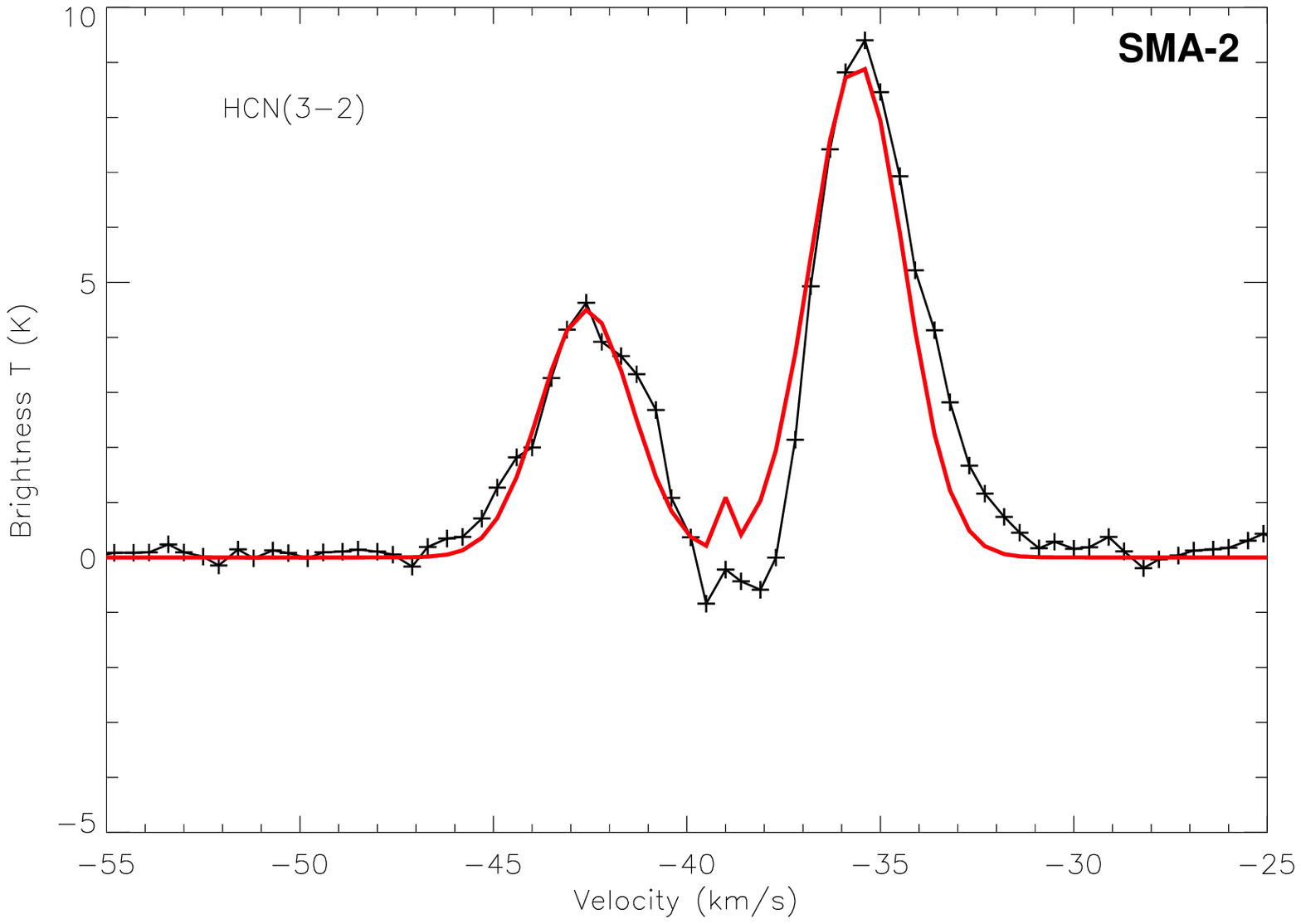}
  \caption{(a) The observed HCN(3-2) spectral profile towards
  SMA-1 (solid black curve) is fitted with the four-layer model
  (red curve) as described by Equation 1, which is discussed in
  Section 4.2.3 The best fitting gives the optical depths and
  excitation temperature for the infall gas in front and rear
  layers ($\tau_{\rm f} = 5.0$, $\tau_{\rm r} = 0.5$,
  $J(T_{\rm f})= 2$ K, $J(T_{\rm r}) = 6$ K), and the outflow
  gas ($\tau_{\rm B} = 0.08$, $\tau_{\rm R} = 0.06$,
  $J(T_{\rm out}) = 76$ K), as well as the dust
  ($\tau_{\rm c}=0.5$, $J(T_{\rm c})f_{\rm c} = 7$ K),
  respectively. The best fitted velocities and velocity
  dispersions are ($V_{\rm in} = 0.9$ km s$^{-1}$,
  $\sigma_{\rm in}$ = 0.4 km s$^{-1}$) and ($V_{\rm out}$ =
  3.6 km s$^{-1}$, $\sigma_{\rm out} = 1.6$ km s$^{-1}$) for
  the infall and the outflow, respectively. The blue dashed and
  dotted lines represent the best fitted spectral components for
  the infall and the outflow described in Equation 2,
  respectively. The line wings are not fitted with the
  four-layer model; instead, they are simply fitted using a
  power-law profile with an index of 2. The black dashed line is
  the residual of subtracting the best fitted model from the
  data. (b) The observed HCN(3-2) spectral profile towards SMA-2
  is fitted to a model with outflow alone ($V_{\rm in}$ = 0
  km~s$^{-1}$) as shown in red curve.}
\end{figure}

\end{document}